\documentclass[conference,a4paper]{IEEEtran}
\IEEEoverridecommandlockouts

\usepackage{cite}
\usepackage{amsmath,amssymb,amsfonts}
\usepackage{graphicx}
\usepackage{textcomp}
\usepackage{xcolor}

\usepackage{booktabs}
\usepackage{makecell}
\usepackage{multirow}
\usepackage[caption=false]{subfig}
\usepackage{adjustbox}

\usepackage{siunitx}
\usepackage{acronym}

\usepackage{svg}
\usepackage{pgfplots}
\pgfplotsset{compat=newest}

\acrodef{ppl}[PPL]{Pulsed Power Load}
\acrodef{sps}[SPS]{Shipboard Power System}
\acrodef{dg}[DG]{Diesel Generator}
\acrodef{bess}[BESS]{Battery Energy Storage System}
\acrodef{ess}[ESS]{Energy Storage System}
\acrodef{afe}[AFE]{Active Front End}
\acrodef{msb}[MSB]{Main Switch Board}
\acrodef{smes}[SMES]{Superconducting Magnetic Energy Storage}
\acrodef{fess}[FESS]{Flywheel Energy Storage System}
\acrodef{avr}[AVR]{Automatic Voltage Regulator}
\acrodef{emt}[EMT]{Electromagnetic Transient}
\acrodef{esr}[ESR]{Equivalent Series Resistance}

\DeclareSIUnit{\kW}{kW}
\DeclareSIUnit{\Wh}{Wh}
\DeclareSIUnit{\MWh}{MWh}
\DeclareSIUnit{\Wp}{Wp}
\DeclareSIUnit{\pu}{p.u.}
\DeclareSIUnit{\EUR}{\mbox{\text{\euro}}}
\DeclareSIUnit{\EURkWh}{\mbox{\text{\euro}}/kWh}
\DeclareSIUnit{\EURkW}{\mbox{\text{\euro}}/kW}
\DeclareSIUnit{\MEUR}{M\mbox{\text{\euro}}}
\DeclareSIUnit{\EURMWh}{\mbox{\text{\euro}}/MWh}
\DeclareSIUnit{\Nm}{Nm}

\begin{document}

\title{Linear Modeling of Hybrid Shipboard Power Systems under Pulsed Power Load Operation}
\author{
\IEEEauthorblockN{
Daniele Roncagliolo,
Pietro Cepollini,
Fabio D'Agostino,
Marco Gallo,
and Federico Silvestro
}

\IEEEauthorblockA{
\textit{DITEN, University of Genova,Genova, Italy}\\
daniele.roncagliolo@edu.unige.it,
pietro.cepollini@edu.unige.it,
fabio.dagostino@unige.it,\\
marco.gallo@edu.unige.it,
federico.silvestro@unige.it
}
}

\maketitle

\begin{abstract}
A model-based approach is a valuable tool in the power system design process, enabling the integration of component models and the evaluation of dynamic interactions. Simplified models support rapid preliminary assessments by allowing parameters and configurations to be adjusted. This paper investigates the effects of high-bandwidth loads in hybrid AC--DC shipboard power systems.
The problem is addressed by developing a linear model aimed at capturing the relevant dynamics over the time scale of interest with reduced computational effort.
A LVDC subsystem has been developed with a battery energy storage system that supports the grid during pulsed power load operation. A control strategy has been proposed to satisfy transient requirements.
The proposed approach is validated under different pulsed power load conditions using a high-fidelity model developed in DIgSILENT PowerFactory, which preserves nonlinearities and includes converter switching effects and digital control implementation.
The validation shows that the model captures the relevant transient behavior with significantly reduced simulation time, while maintaining acceptable accuracy, making it suitable for rapid preliminary design studies.
\end{abstract}


\section{Introduction}
\begin{figure}[!t]
 \centering
 \includegraphics[width=0.8\columnwidth]{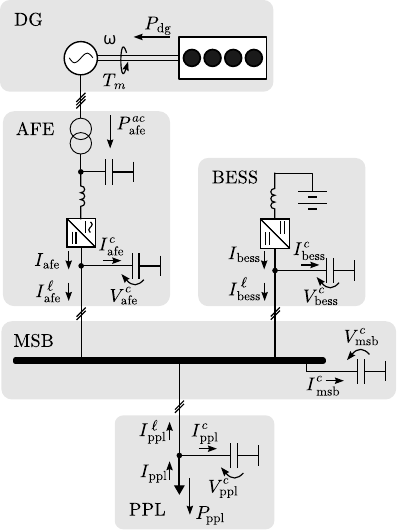}
 \caption{System single-line diagram.}
 \label{fig:SingleLine}
\end{figure}
The growing integration of high-bandwidth loads, such as \acp{ppl}, and energy storage systems is playing a significant role in the ongoing transition toward hybrid AC--DC \ac{sps}. In particular, \acp{ppl} pose significant challenges due to their rapid and substantial changes in power demand, affecting the power quality of both DC and AC subsystems.
Military standards, such as STANAG-1008 \cite{stanag}, define \acp{sps} requirements in terms of voltage and frequency deviations in response to rapid changes in load, making transient performance a key aspect of system design. 

Several \acp{ess} have been proposed in the literature as a means of mitigating the disturbances introduced by \acp{ppl} \cite{ESSoverview}. Considering the DC nature of most \acp{ess} and \acp{ppl}, together with the increasing interest in on-board LVDC distribution systems, the adoption of DC sections represents a promising design solution for future \acp{sps} \cite{Zonallvdc}.

In the case of DC \acp{ppl} in hybrid \acp{sps}, rapid power variations cause DC voltage deviations and, consequently, a response from the voltage control loops to modulate the current contribution of the sources. The resulting converter current variations affect the AC-side power balance, producing voltage and frequency deviations to which the \acp{dg} respond through mechanical torque and excitation control, respectively.

In this context, IEEE Std. 45.1-2023 \cite{ieee45.1} provides design guidelines for shipboard systems with voltage levels up to \SI{13}{\kilo\volt} AC and \SI{1}{\kilo\volt} DC, and defines stringent limits on frequency, AC-bus and DC-bus voltage deviations during \ac{ppl} operation.

Given the strict regulatory constraints and the complexity of such systems the preliminary system design requires effective model-based tools for transient assessment and compliance verification. Detailed modeling approaches, such as those proposed in \cite{EVH_Detailed} for electric vehicles and in \cite{SPS_detailed_model3} for hybrid \acp{sps}, address the analysis of dynamic systems through detailed mathematical formulations and numerical simulations. Detailed models provide high-fidelity representations of power systems, but their computational burden can limit their use in iterative design and sensitivity studies. Simplified modeling approaches have therefore been proposed as effective tools to capture transient dynamics with less computational effort \cite{PW_Linearization_nonlinearModels}. In the context of \acp{sps}, simplified models have also been investigated as computationally efficient tools for transient studies \cite{SPS_reduced_model1,SPS_reduced_model2}. These approaches rely either on reduced order and average value models, which still retain non-linear algebraic relationships, or on modular non-linear time-domain formulations, such as bond graphs, which require step-by-step numerical integration.

However, these formulations still require numerical solvers, limiting the achievable computational efficiency. This becomes particularly relevant when analysing multiple operating conditions and fast \ac{ppl} transients. To overcome this limitation, this work proposes a fully linearised analytical model of a hybrid AC--DC power system for the assessment of transients caused by \ac{ppl} operation. The proposed formulation avoids iterative numerical integration and the solution of non-linear algebraic equations, allowing the system dynamics to be solved in closed form. The model is linearised with respect to voltage and frequency, assuming small deviations from their nominal values, consistently with the transient limits specified by naval reference standards.

The study is carried out on a representative feeder derived from the \ac{sps} model presented in \cite{pplWithSMESS_interno}, including an AC generation system, an interface converter, and an LVDC section integrating a \ac{ppl} and a \ac{bess}. The model focuses on the system-level dynamics of interest, neglecting faster switching and inner current control dynamics. This enables rapid validation of complex hybrid \ac{sps} designs under different loading conditions, without resorting to computationally intensive detailed simulations.

A benchmark switchin EMT model is developed in the DIgSILENT PowerFactory environment, following the approach presented in \cite{pplWithSMESS_interno}, to provide a high-fidelity reference for the proposed model. Validation is performed for four \ac{ppl} loading conditions to assess the fidelity of the model in reproducing the dynamics of interest over different disturbance magnitudes.

The validation results show that the linear model reproduces the transient behavior of the considered quantities within the considered conditions. It also enables a preliminary system performance assessment based on compliance with the voltage and frequency deviation limits specified by the standards, and with the nominal values for the battery-side current and the alternator shaft torque.

\section{System Description}

The considered \ac{sps} is derived from a notional microgrid, including \SI{6.6}{\kilo\volt} AC busbars, four \SI{2.65}{\mega{\volt\ampere}} \acp{dg}, aggregated loads, propulsion loads, and an \ac{afe} connected through a \SI{6.6}{\kilo\volt}/\SI{0.44}{\kilo\volt} transformer \cite{pplWithSMESS_interno}. A \ac{ppl} is integrated into a dedicated LVDC section with a nominal voltage $V^\mathrm{dc}_\mathrm{n}$ of \SI{1}{\kilo\volt}, together with a \ac{bess} designed to provide fast DC voltage regulation service.

Figure \ref{fig:SingleLine} shows the single-line diagram of the representative feeder considered in this work, including the main network components that affect the dynamic behavior of the system during \ac{ppl} operation.

The \ac{bess} consists of a \SI{650}{\volt} battery, interfaced through a bidirectional DC-DC boost converter with a continuous power rating of \SI{1}{\mega\watt}, equipped with an inductive filter ($L_{\mathrm{bess}}=\SI{370}{\micro\henry}$) and capacitive filters designed according to \cite{boost_filters_design}.
The \ac{afe} is equipped with an AC-side LCL filter, featuring a converter-side inductance $L_{\mathrm{afe}} = \SI{25}{\micro\henry}$ and a filter capacitance $C_{\mathrm{afe}} = \SI{820}{\micro\farad}$, whose grid-side inductance $L_{\mathrm{transf}}$ is provided by the interface transformer \cite{AFE_LCL_1,AFE_LCL_2}. The DC-side bus capacitor is designed according to \cite{AFE_dc-bus}.
The \ac{bess} and  \ac{ppl} are connected to the \ac{msb} through \SI{10}{\meter} cables, while the \ac{afe} is connected via a \SI{1}{\meter} cable. All cables are assumed to be equal with a resistance per unit length of $R_{\ell}=\SI{1}{\milli\ohm\per\meter}$ and an RL time constant of $\tau^{\mathrm{rl}}=\SI{1}{\milli\second}$.
Table \ref{tab:capacitance_parameters} summarises the LVDC capacitances and their respective time constants, \textit{i.e.,} the product of the capacitance and the \ac{esr}.

\begin{table}[t]
    \centering
    \caption{DC-bus capacitor parameters and ESR time constants.}
    \label{tab:capacitance_parameters}
    \renewcommand{\arraystretch}{1.2}
    \setlength{\tabcolsep}{3.5pt}
    \begin{tabular}{cc cc cc cc}
        \toprule
        \multicolumn{2}{c}{AFE} & \multicolumn{2}{c}{BESS} & \multicolumn{2}{c}{PPL} & \multicolumn{2}{c}{MSB} \\
        \cmidrule(lr){1-2} \cmidrule(lr){3-4} \cmidrule(lr){5-6} \cmidrule(lr){7-8}
        $C_{\mathrm{afe}}$ & $\tau^{\mathrm{esr}}_{\mathrm{afe}}$ & $C_{\mathrm{bess}}$ & $\tau^{\mathrm{esr}}_{\mathrm{bess}}$ & $C_{\mathrm{ppl}}$ & $\tau^{\mathrm{esr}}_{\mathrm{ppl}}$ & $C_{\mathrm{msb}}$ & $\tau^{\mathrm{esr}}_{\mathrm{msb}}$ \\
        \midrule
        \SI{5.8}{\milli\farad} & \SI{5.8}{\micro\second} & \SI{44}{\milli\farad} & \SI{440}{\micro\second} & \SI{5}{\milli\farad} & \SI{5}{\micro\second} & \SI{50}{\milli\farad} & \SI{500}{\micro\second} \\
        \bottomrule
    \end{tabular}
\end{table}
During \ac{ppl} operation, the transient performance is determined by the interaction between the \ac{bess} and \ac{afe} control systems and the passive dynamics of the LVDC section. The control strategy objective is to make the \ac{bess} supply the fast transient power required during \ac{ppl} operation, while the \ac{afe} progressively provides the steady-state power contribution from the AC generation system.

The control systems of the \ac{bess} and the \ac{afe} are organised according to a hierarchical structure composed of an inner current control loop and an outer DC voltage control loop. The \ac{bess} control structure is based on the approach proposed in \cite{imperix_cascaded_voltage_control}. A load-current feedforward term is included to improve the dynamic response of DC voltage regulation by anticipating the current demand imposed by the \ac{ppl}. The \ac{afe} adopts a grid-following control structure without a feedforward contribution \cite{imperix_afe_control}. The current and voltage control loops are tuned according to the modulus optimum and symmetrical optimum criteria, respectively \cite{SO_MO}.

The proposed control strategy adopts DC voltage droop regulation for the \ac{bess}, based on the output line current $I_{\mathrm{bess}}^{\ell}$, with the voltage reference computed as: \begin{equation} V_{\mathrm{bess}}^{c\,*} = V_{0}^{c} - K_{\mathrm{b}}\, I_{\mathrm{bess}}^{\ell}, \label{eq:dc_droop} \end{equation} where $K_{\mathrm{b}}$ is the droop coefficient and $V_{0}^{c}$ is the no-load voltage reference, assumed equal to $V^\mathrm{dc}_\mathrm{n}$. The \ac{afe} voltage reference is instead set equal to the nominal \ac{msb} voltage, $V^\mathrm{dc}_\mathrm{n}$.

When the \ac{ppl} increases its power demand, the resulting increase in DC current demand is initially compensated by the \ac{bess}, whose load-current feedforward allows for a faster response to the disturbance. The droop mechanism enables the BESS to regulate its power contribution during the transient phase. The \ac{afe} voltage control loop acts as a slower secondary regulation reserve, which restores the voltage, progressively reducing the \ac{bess} current contribution.

As a result, the control structure defines a dynamic power-sharing mechanism in which the \ac{bess} provides the fastest response during \ac{ppl} transients, while the \ac{afe} progressively takes over the steady-state power contribution.

\section{System Modeling}

The model aims to capture the dominant transient dynamics during \ac{ppl} operation with limited computational effort. The study focuses on the coupled behavior of frequency and DC voltage, which are driven by the electromechanical power imbalance on the AC-side and by the nodal current imbalance within the LVDC section. For this purpose, the formulation relies on three main assumptions.

i) The model neglects the electromagnetic phenomena associated with the commutation of converters.

ii) Inner current control loops and measurement filters are considered at steady-state.

iii) Power variations due to changes in frequency or DC voltage are assumed to be negligible compared to those caused by torque and current variations, respectively.

This condition is consistent with the operating range of practical interest for \ac{sps} transient assessment, since the reference standards specify allowable limits for voltage and frequency deviations. Therefore, within the range defined by these requirements, the linear model is expected to provide an adequate representation of the system dynamics while maintaining a significantly lower computational effort than detailed EMT models.
In the following model formulation, all quantities are expressed in \si{\pu}.


In terms of small-signal variations, the \ac{dg} mechanical power $p_{\mathrm{dg}}$ is a function of mechanical speed $\omega$ and torque $t_m$ is expressed as:
\begin{equation}
   \Delta p_{\mathrm{dg}} = \omega_0\Delta t_m+ t_{m0}\Delta\omega, 
\end{equation}
where $\omega_0$ is the initial mechanical speed and $t_{m0}$ is the initial mechanical torque. Therefore, assuming small speed variations around nominal frequency, it follows that:
\begin{equation}
\Delta p_{\mathrm{dg}} \approx \Delta t_m.
\label{eq:PTm_lin_dg}
\end{equation}

Similarly, the linearised relationship between \ac{afe} active power $p_{\mathrm{afe}}^{ac}$ with DC voltage $v_{\mathrm{afe}}^{c}$ and current $i_{\mathrm{afe}}^{\ell}$ is derived as:
\begin{equation}
\Delta p_{\mathrm{afe}}^{ac}
=
i_{\mathrm{afe0}}^{\ell} \Delta v_{\mathrm{afe}}^{c} +v_{\mathrm{afe0}}^{c} \Delta i_{\mathrm{afe}}^{\ell},
\label{eq:afe_power_current}
\end{equation}
where $i_{\mathrm{afe0}}^{\ell}$ and $v_{\mathrm{afe0}}^{c}$ denote the steady-state initial conditions. Consequently, assuming negligible DC voltage variations around the nominal point:
\begin{equation}
\Delta p_{\mathrm{afe}}^{ac}
\approx
\Delta i_{\mathrm{afe}}^{\ell}.
\label{eq:PI_lin_afe}
\end{equation}

Following the same approach, the \ac{ppl} current under small deviations reduces to:
\begin{equation}
\Delta i_{\mathrm{ppl}}
\approx
-\Delta p_{\mathrm{ppl}},
\label{eq:PI_lin_ppl}
\end{equation}
where $i_{\mathrm{ppl}}$ is the \ac{ppl} injected current and $p_{\mathrm{ppl}}$ is the absorbed power.



\subsection{DC Voltage Dynamics}

The DC voltage dynamics are governed by the interaction between the local capacitive branches and the resistive-inductive lines connecting each DC element to the \ac{msb}.
The local capacitor branch impedance characterises the response of the generic $i$-th capacitor current $i_i^\mathrm{c}$ and voltage $v^{\mathrm{c}}_i$:
\begin{equation}
Z_i^{\mathrm{c}}(s)
=
\frac{\Delta v^{\mathrm{c}}_i(s)}{\Delta i_i^\mathrm{c}(s)}
=
\frac{1+s\,\tau_i^{\mathrm{rc}}}{s c_i},
\label{eq:dcbus_tf}
\end{equation}
where $c_i$ and $\tau_i^{\mathrm{rc}}$ are the local capacitance and its time constant, respectively.

Furthermore, the line admittance defines the line current $i_i^{\ell}$ driven by the voltage difference between the \ac{msb} and the $i$-th element:
\begin{equation}
Y_i^{\ell}(s)
=
\frac{\Delta i_i^{\ell}(s)}
{\Delta v^{\mathrm{c}}_{\mathrm{msb}}(s) - \Delta v^{\mathrm{c}}_i(s)}
=
\frac{1}{2\,r\,\ell_i}
\frac{1}{1+s\,\tau^{\mathrm{rl}}},
\label{eq:dcline_tf}
\end{equation}
where $r$ and $\ell_i$ denote the per-unit line resistance and the connection length of the $i$-th element to the \ac{msb}, respectively.

Finally, the dynamical relationship between the \ac{msb} capacitor current $i_{\mathrm{msb}}^{\mathrm{c}}$ and its voltage $v^{\mathrm{c}}_{\mathrm{msb}}$ can be derived as:
\begin{equation}
\Delta v_{\mathrm{msb}}^{\mathrm{c}}(s) 
= 
Z^{c}_{\mathrm{msb}}(s)\,\Delta i_{\mathrm{msb}}^{\mathrm{c}}(s),
\label{eq:dcbus_io}
\end{equation}
where the total capacitor current satisfies Kirchhoff's current law:
\begin{equation}
\Delta i_{\mathrm{msb}}^{\mathrm{c}}(s)
=
\sum_i \Delta i_{i}^{\ell}(s),
\label{eq:MSB_Current_Balance}
\end{equation}
with $i_{i}^{\ell}$ representing the $i$-th line current flowing into the \ac{msb}.

\subsection{Frequency Dynamics}

The AC system frequency dynamics are dictated by the relationship between the generator speed deviation $\Delta \omega$ and the balance between the generated power $\Delta p_{\mathrm{dg}}$ and the \ac{afe} absorbed power $\Delta p_{afe}$.
Assuming a lossless inverter and neglecting the friction factor, the relationship is described by the swing equation:
\begin{equation}
\Delta \omega(s) = \frac{1}{2Hs}\,[\Delta p_{\mathrm{dg}}(s)-\Delta p_{\mathrm{afe}}(s)],
\label{eq:swing_io}
\end{equation}
with $H$ being the inertia constant of the \ac{dg}.

\subsection{Diesel Genset}
The \ac{dg} dynamics are described by the following relationship:

\begin{equation}
\Delta t_m(s) = G_{\mathrm{dg}}(s)\,\Delta \omega(s) . 
\label{eq:dg_StateEq}
\end{equation}

Considering the combustion delay $T_d$, the resulting transfer function can be written as:

\begin{equation}
G_{\mathrm{dg}}(s) = \frac{E(s)A(s)}{1 + k_{\mathrm{dg}}\,E(s)A(s)}\,\cdot\,e^{-sT_d},
\label{eq:dg_tf}
\end{equation}
where $A(s)$ and $E(s)$ represent the actuator and the electric control box transfer function related to the governor system, respectively \cite{PSSE_ModelLibrary_v32}. $k_{\mathrm{dg}}$ is the droop coefficient.
\subsection{Active Front End}
The \ac{afe}
is modeled through the associated admittance seen at the \ac{msb} $Y_{\mathrm{afe}}(s)$. Therefore, the \ac{afe} line current variation is expressed as:
\begin{equation}
 \Delta i_{\mathrm{afe}}^{\ell}(s)
 =
 \,-Y_{\mathrm{afe}}(s)\,\Delta v_{\mathrm{msb}}^\mathrm{c}(s)\,,
\label{eq:afe_StateEq}
\end{equation}
where:
\begin{equation}
 Y_{\mathrm{afe}}(s)
 =
 \frac{
 Y_{\mathrm{afe}}^{\ell}(s)
 \left[
 1+Z_{\mathrm{afe}}^{c}(s)G_{\mathrm{afe}}^{pi}(s)
 \right]}
 {1+Z_{\mathrm{afe}}^{c}(s)
 \left[
 G_{\mathrm{afe}}^{pi}(s)+Y_{\mathrm{afe}}^{\ell}(s)
 \right]},
\label{eq:g_afe}
\end{equation}
where $G_{\mathrm{afe}}^{pi}(s)$ represents the voltage controller transfer function of the \ac{afe}.

\subsection{Battery Energy Storage System}

The \ac{bess} is described by the admittance seen at the \ac{msb}. Therefore, the current variation of the \ac{bess} line is expressed as:

\begin{equation}
\Delta i_{\mathrm{bess}}^{\mathrm{\ell}}(s) =-\, Y_{\mathrm{bess}}(s)\,\Delta v^{\mathrm{c}}_{\mathrm{msb}}(s), 
\label{eq:bess_StateEq}
\end{equation}
with the \ac{bess} admittance:
\begin{equation}
Y_{\mathrm{bess}}(s) =
\frac{Y_{\mathrm{bess}}^{\ell}(s)}
{1 + k_{\mathrm{b}}\,Y_{\mathrm{bess}}^{\ell}(s)\,G_{\mathrm{b}}(s)}\,,
\label{eq:bess_tf}
\end{equation}
where $k_{\mathrm{b}}$ is the per-unit droop coefficient of the voltage regulator, and $G_{\mathrm{b}}(s)$ describes the closed-loop dynamics of the converter control system, given by:

\begin{equation}
G_{\mathrm{b}}(s) =
\frac{G_{\mathrm{bess}}^{pi}(s)\,Z_{\mathrm{bess}}^{c}(s)}
{1 + G_{\mathrm{bess}}^{pi}(s)\,Z_{\mathrm{bess}}^{c}(s)}\,,
\label{eq:bess_control}
\end{equation}
where $G_{\mathrm{bess}}^{pi}(s)$ represents the voltage controller transfer function of the \ac{bess}.
\subsection{Pulsed Power Load}
The \ac{ppl} line current depends on the \ac{msb} voltage $v_{\mathrm{msb}}^\mathrm{c}$ and \ac{ppl} current setpoint $i_{\mathrm{ppl}}$. Specifically, the voltage variation dependence is described by $Y_{\mathrm{ppl}}(s)$, whereas the current setpoint dependence is modeled with $G_{\mathrm{ppl}}(s)$.
Therefore, the \ac{ppl} line current variation is expressed as:
\begin{equation}
 \Delta i_{\mathrm{ppl}}^{\ell}(s)
 =
 \,-Y_{\mathrm{ppl}}(s)\,\Delta v_{\mathrm{msb}}^\mathrm{c}(s)
 +
 G_{\mathrm{ppl}}(s)\,\Delta i_{\mathrm{ppl}}(s)\,,
\label{eq:PPL_StateEq}
\end{equation}
$\Delta i_{\mathrm{ppl}}$ is derived from \eqref{eq:PI_lin_ppl} and the aforementioned transfer functions are derived as:
\begin{equation}
 G_{\mathrm{ppl}}(s)
 =
 \frac{Y_{\mathrm{ppl}}^{\ell}(s)\,Z_{\mathrm{ppl}}^{c}(s)}
 {1+Y_{\mathrm{ppl}}^{\ell}(s)\,Z_{\mathrm{ppl}}^{c}(s)}\,,
\label{eq:Y_ppl}
\end{equation}
and:
\begin{equation}
 Y_{\mathrm{ppl}}(s)
 =
 \frac{Y_{\mathrm{ppl}}^{\ell}(s)}
 {1+Y_{\mathrm{ppl}}^{\ell}(s)\,Z_{\mathrm{ppl}}^{c}(s)} \,.
\label{eq:g_ppl}
\end{equation}

\section{Simulation and discussion}

A switching EMT model is developed in DIgSILENT PowerFactory, consistently with the approach presented in \cite{pplWithSMESS_interno}. The LVDC subsystem is represented as a two-wire distribution system, with lines modeled as RL branches.
The \ac{afe} and DCDC boost converters are implemented using switching functions. Cascaded voltage and current control loops are included in detail considering digital control implementation effects.
The \ac{ppl} is implemented as a power controlled DC current source.
The \ac{ppl} power profile is evaluated under different loading levels, specifically at \SI{100}{\percent}, \SI{200}{\percent}, \SI{300}{\percent}, and \SI{400}{\percent} scaling of the original profile presented in \cite{MIL_STD_1399_Part_1,MIL_STD_1399_Part_2}.

\begin{figure}[!t]
 \centering
 \includegraphics[width=\columnwidth]{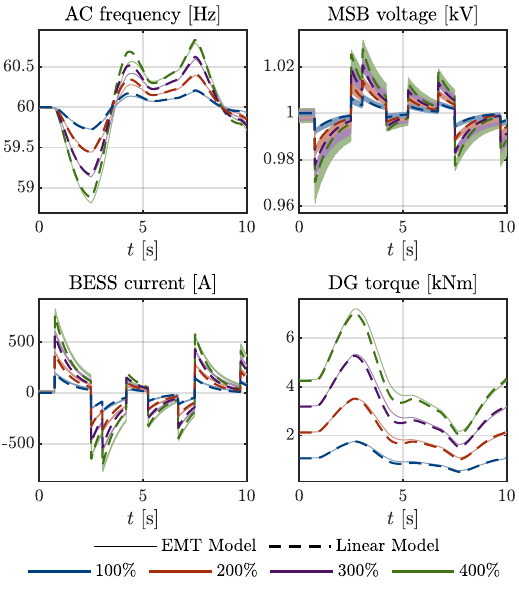}
 \caption{EMT and linear models transient responses for different \ac{ppl} scalings.}
 \label{fig:validation}
\end{figure}

Figure~\ref{fig:validation} compares the linear and EMT models in terms of frequency, \ac{msb} voltage, mechanical torque and battery current transients across the different loading levels. 
The system transient performance is assessed through the indicators reported in Table \ref{tab:performance_assessment}. Frequency and DC voltage modulations, denoted as $f.m.$ and $v.m.$, are expressed in percentage, while battery current and \ac{dg} torque peaks, $i_{\mathrm{bess}}^{\max}$ and $t_m^{\max}$, are expressed in \si{\pu} with respect to their nominal values.
\begin{table}[t]
\centering
\caption{Transient Performance Indicators}
\label{tab:performance_assessment}
\renewcommand{\arraystretch}{1.25}
\scriptsize
\begin{tabular}{ccccc}
\toprule
\textbf{PPL Scaling} &
\makecell{\textbf{$\mathrm{f.m.}$ [\%]}\\EMT / Lin.} &
\makecell{\textbf{$\mathrm{v.m.}$ [\%]}\\EMT / Lin.} &
\makecell{\textbf{$i_{\mathrm{bess}}^{\max}$ [\si{\pu}]}\\EMT / Lin.} &
\makecell{\textbf{$t_m^{\max}$ [\si{\pu}]}\\EMT / Lin.} \\
\midrule
\SI{100}{\percent} & $0.40 / 0.41$ & $0.86 / 0.72$ & $0.16 / 0.15$ & $0.16 / 0.16$ \\
\SI{200}{\percent} & $0.79 / 0.81$ & $1.70 / 1.44$ & $0.32 / 0.30$ & $0.32 / 0.32$ \\
\SI{300}{\percent} & $1.21 / 1.22$ & $2.59 / 2.16$ & $0.49 / 0.46$ & $0.48 / 0.47$ \\
\SI{400}{\percent} & $1.66 / 1.63$ & $3.51 / 2.87$ & $0.66 / 0.61$ & $0.65 / 0.63$ \\
\bottomrule
\end{tabular}
\end{table}

Frequency modulation exceeds the admissible \SI{0.5}{\percent} limit for \ac{ppl} scaling levels equal to \SI{300}{\percent} and \SI{400}{\percent}, whereas the DC voltage modulation remains below the \SI{5}{\percent} limit over the considered range. The battery current and \ac{dg} torque peaks also remain below their nominal values, indicating that no-source operating constraints are exceeded in the analysed cases.

\begin{figure}[!t]
\centering
\includegraphics[width=\columnwidth]{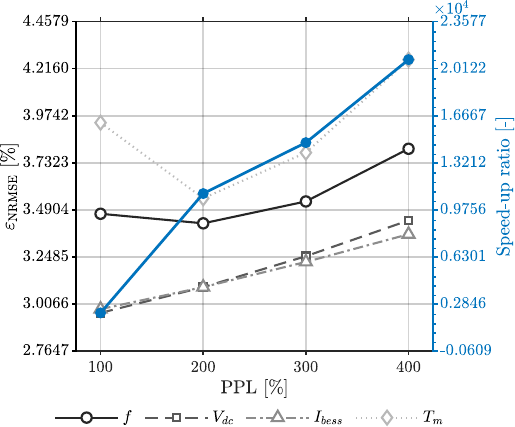}
\caption{Accuracy and computational speed-up of the linear model for different \ac{ppl} scaling levels.}
\label{fig:3D}
\end{figure}

To further quantify the agreement between the two models, the Normalised Root Mean Square Error (NRMSE) is evaluated over the transient window, using the EMT response as reference and its peak-to-peak range for normalisation:
\begin{equation}
    \mathrm{NRMSE} =
\frac{\mathrm{RMS}(x_{\mathrm{lin}}-x_{\mathrm{EMT}})}
{x_{\mathrm{EMT}}^{\max}-x_{\mathrm{EMT}}^{\min}}\times100.
\end{equation}
The simulation time speed-up ratio is also considered to assess the computational advantage of the linear model. Defining the speed-up ratio as the ratio of the EMT simulation time to that of the linearised model, Figure~\ref{fig:3D} reports the NRMSE and the corresponding speed-up ratio for the considered \ac{ppl} scaling levels. As expected, the discrepancy between the two models increases with the \ac{ppl} scaling, consistently with the small-signal assumptions adopted in the model formulation. Nevertheless, the linear model maintains a good agreement with the EMT response while providing a significant reduction in simulation time, with a speed-up approximately four orders of magnitude lower than the detailed model.

For operating conditions within the transient limits specified by the reference standards, the adopted assumptions remain consistent with the expected range of voltage and frequency deviations. The results also show that the linear model maintains acceptable accuracy for larger \ac{ppl} transients beyond these limits. these results support the validity of the linearised formulation for preliminary dynamic assessments. The achieved reduction in computational time is particularly relevant for the rapid evaluation of multiple operating conditions, sensitivity analyses, and critical \ac{ppl} transient scenarios, where the computational cost of detailed EMT simulations can become prohibitive.
\section{Conclusions}






This paper presents a model-based approach for the preliminary transient assessment of a hybrid AC--DC \ac{sps} supplying a high bandwidth load. A linear model is developed to capture the relevant transient dynamics with low computational effort and to support preliminary assessment of design specifications and standard limits.

The LVDC subsystem includes a \ac{bess} that provides fast support during \ac{ppl} operation, while the \ac{afe} ensures slower DC voltage regulation and transfers the steady-state power demand to the AC-side generation system. This control strategy enables transient power sharing with the storage system and limits the impact of rapid DC-side load variations on the AC subsystem.

The proposed approach is validated under four \ac{ppl} loading conditions using a high-fidelity switching EMT model developed in DIgSILENT PowerFactory. The results show that the linear model reproduces the relevant transient behavior and provides preliminary performance indicators, including frequency and DC voltage modulations, \ac{dg} torque, and battery current peaks. The comparison also demonstrates a significant reduction in simulation time, reaching up to four orders of magnitude with respect to the detailed EMT model, while maintaining acceptable accuracy even for \ac{ppl} conditions exceeding the considered standard limits.

Overall, the proposed model provides an effective balance between computational efficiency and accuracy, making it suitable for supporting rapid preliminary design studies, control development, sensitivity analyses, and compliance assessment of hybrid AC--DC \acp{sps} subject to \ac{ppl} operation.

Future work will extend the formulation by including AC voltage and reactive power dynamics and by considering a broader range of operating conditions.For larger \ac{ppl} transients or operating conditions farther from the linearisation point, a piecewise linear formulation could be investigated to further improve model accuracy.


\end{document}